\documentclass[12pt]{article}

\usepackage{amsmath}
\usepackage{amssymb}
\usepackage{graphicx}

\newcommand{\sign}{\mathop{\mathrm{sign}}}

\def\PL{ Phys. Lett. }
\def\PR{ Phys. Rev. }
\def\PRL{ Phys. Rev. Lett. }
\def\APJ{ Astroph.~J.~}
\def\AJ{ Astron.~J.~}

\begin{document}


\title{Exactly Solvable SFT Inspired Phantom Model}

\author{
I.~Ya.~Aref'eva\footnote{\texttt{arefeva@mi.ras.ru}, Steklov Mathematical
Institute, Russian Academy of Sciences},
A.~S.~Koshelev\footnote{\texttt{koshelev@mi.ras.ru}, Steklov Mathematical
Institute, Russian Academy of Sciences}
\\and\\
S.~Yu.~Vernov\footnote{\texttt{svernov@theory.sinp.msu.ru}, Skobeltsyn
Institute of Nuclear Physics, Moscow State University}}

\date{~}

\maketitle

\begin{abstract}
An exact solution to the Friedmann equations with a string inspired
phantom scalar matter field is constructed and the absence of the
``Big Rip'' singularity is shown explicitly. The notable features of
the concerned model are a ghost sign of the kinetic term and a
special polynomial form of the effective tachyon potential. The
constructed solution is stable with respect to small fluctuations of
the initial conditions and special deviations of the form of the
potential.
\end{abstract}

\section{INTRODUCTION}

The combined analysis of the type Ia supernovae, galaxy clusters
measure\-ments and WMAP (Wilkin\-son Microwave Anisotropy Probe)
data brings out clearly an evidence of the accelerated expansion of
the Universe~\cite{Perlm}-\cite{Spergel}. The cosmologi\-cal
acceleration strongly indicates that the present day Universe is
dominated by a smoothly distributed slowly varying cosmic fluid with
a negative pressure, the so-called dark
energy~\cite{S-St}-\cite{CaldwellPW} (alternative approaches are
presented, for example, in~\cite{CarTrod}).

To specify different types of cosmic fluids one usually uses a
phenomenological relation between the pressure density $p$ and the
energy density $\varrho$, corresponding to each component of fluid
\begin{equation*}
 p=w\varrho,
\end{equation*}
where $w$ is the equation-of-state parameter or, for short, the state
parameter. A component with negative $w$ corresponds to the dark energy.
Contemporary experiments, including WMAP, give strong support that at present
time the dark energy state parameter is close to $-1$~\cite{Spergel},
\cite{knop}-\cite{Tonry}. In particular,  it follows from the current
observational bounds~\cite{Riess1} that $w$ belongs to the interval:
\begin{equation*}
-1.46 < w <-0.78. \label{wdata}
\end{equation*}
at the $95\%$ confidence level.

 From the theoretical point of view the
above-mentioned domain of $w$ covers three essentially different cases:
$w>-1,~w=-1$ and $w<-1$.

\begin{itemize}
\item The first case, $w >-1$, is achieved in quintessence
models~\cite{Wetterich}-\cite{Sami}, which are cosmological models with a
scalar field. Such types of models are quite acceptable, but there is a
question of an origin of the scalar field. To comply with astronomical
experimental data this scalar field should be extra light and hence it does
not belong to the Standard Model set of fields~\cite{Okun}.

\item The second case, $w=-1$, is realized by means of the cosmological
constant. This scenario is admissible from a general point of view
except for a problem of an order of the magnitude of the
cosmological constant. It should be $10^{120}$ times less than the
natural theoretical prediction~\cite{CaldwellPW}.

\item The third case, $w<-1$, is called a "phantom" one and can be realized
due to a scalar field with a ghost (phantom) kinetic term. In this
case all natural energy conditions are violated and there are
problems of instability at classical and quantum
levels~\cite{Caldwell}-\cite{Arefeva}. Since experimental data do
not contradict with a possibility of $w<-1$ and moreover a direct
search strategy to test the inequality $w<-1$ has been
proposed~\cite{0312430}, it is interesting to find a
noncontradictory model with the condition $w<-1$
satisfied\footnote{Let's note the model~\cite{nemanja}, in which
$w<-1$ is a result of axion consideration.}.
\end{itemize}

Note the $\kappa$-essence models~\cite{mukhanov} can have both $w<-1$ and
$w\geqslant -1$. At the same time a dynamical transition from the region
$w\geqslant -1$ to the region $w<-1$ or vice versa is forbidden under general
assumptions~\cite{Vikman} and is possible only under special
conditions~\cite{andrianov}.

Let us recall that in models with a constant state parameter $w$,
less than $-1$, and the spatially flat Friedmann metric the scale
factor tends to infinity and therefore the Universe breaks down at a
finite moment of time. This problem is known as the ``Big
Rip''~\cite{Caldwell03}, see also \cite{Greek_promise}. The simplest
way to overcome this difficulty in models with $w<-1$ is to consider
a scalar field $\phi$ with a negative time component in the kinetic
term~\cite{Melchiorri03,carroll03}. The appearance of a such summand
is possible due to quantum
effects~\cite{StarobinskyYokoyama,Woodard}. However, all proposed
models insist of the instability problem.

A possible way to evade the instability problem for models with
$w<-1$ is to yield a phantom model as an effective one, arising from
a more fundamental theory without a negative kinetic term. In
particular, if we consider a model with higher derivatives such as
$\phi e^{-\Box}\phi$, then in the simplest approximation: $\phi
e^{-\Box}\phi\simeq\phi^2-\phi\Box\phi$, such a model gives a
kinetic term with a wrong (ghost) sign. It turns out, that such a
possibility does appear in the string field theory
framework~\cite{Arefeva}. Since the concerned model is a string
theory approximation,  all stability problems related to a model
with higher derivatives are discarded.

Our goal in the present paper is to construct an analytic solution
in a polynomial model, which is close in some approximation to a
model arising in the string theory, namely in the theory of a
fermionic NSR string with GSO$-$ sector. The scalar field $\phi$ is
an open string theory tachyon, which, according to the Sen's
conjecture~\cite{Sen-g}, describes brane decay, at which a slow
transition to the stable vacuum, correlating with states of the
closed string, takes place. Other dark energy models, which use the
brane-world scenarios, are presented in~\cite{brane}.

The notable feature of the model is a ghost sign of the kinetic term and a
special form of the potential:
\begin{equation}
V(\phi)= \frac12 \left(1-\phi^2\right)^2 +
\frac{1}{12m_p^2} \phi^2\left(3-\phi^2\right)^2.
\label{Vphi}
\end{equation}

In the first approximation  only the first summand in the right hand side of
(\ref{Vphi}) can be obtained from the open string field theory in a flat
metric~\cite{NPB} (which corresponds to $m_p=\infty$). The second summand
allows us to save the form of the interpolating solution obtained in the flat
metric for the case of the Friedmann Universe, i.e. at an arbitrary $m_p$.
Note, that the appearance of higher powers of $\phi$ is possible if we take into
account the heavier  perturbations of the string~\cite{review-sft}, but we can not expect
to obtain  the explicit values of the coefficients of the
potential (\ref{Vphi}) from the string theory. The performed in this paper stability analysis of
solutions points out that the constructed solution is an attractive one and there are
no essential changes in the behavior of the solutions for some variations of
the potential parameters. This gives reasons to assume, that the qualitative behavior of
the physical parameters $H, w, q$ in the model with the potential (\ref{Vphi}) under
certain conditions becomes close to the behavior of the corresponding ones in
the model with a string theory potential.

The constructed solution has no singular points and becomes the de-Sitter
solution at large time. It does not contradict to the general theorems of the
relativity theory because the energy conditions are violated in the considered
model (this question is discussed in~\cite{MacInnes}).

The paper is organized as follows. In Section 2 we describe the
cosmological model, which is an approximation for the string
inspired model, and write down the solutions to the Friedmann equations
with the special polynomial potential. In Section 3 we consider the
phantom dynamics in detail and find out that our solution describes an
accelerating Universe. In Section 4 we study the stability of the
solution with respect to variations of initial data and peculiar
variations of a form of the potential. In Section 5 we do sum up the
obtained results and point out the ways for generalization of the
considered model.

\section{MODEL}


\subsection{Equations of motion}

Let us consider the gravitational model with a phantom scalar field and the
spatially flat Friedmann metric. Since the origin of the phantom field,
considering in this paper, is connected with the string field theory we include
the typical string mass $M_s$ and a dimensionless open string coupling
constant $g_o$ in the action. The action is:
\begin{equation}
 S=\int d^4x \sqrt{-g}\left(\frac{M_p^2}{2M_s^2}R+\frac1{g_o^2}\left(
 +\frac{1}{2}g^{\mu\nu}\partial_{\mu}\phi\partial_{\nu}\phi
-V(\phi)\right)\right), \label{action}
\end{equation}
where $M_p$ is the Planck mass, $g$ is a spatially flat Friedmann metric:
\begin{equation*}
ds^2={}-dt^2+a^2(t)\left(dx_1^2+dx_2^2+dx_3^2\right),
\end{equation*}
where $a(t)$ is the scale factor. The coordinates $(t,x_i)$ and the field $\phi$ are
dimensionless. Hereafter we use the dimensionless parameter
$m_p^2={g_o^2M_p^2}/{M_s^2}$.

Independent equations of motion are as follows
\begin{equation}
3H^2=\frac{1}{m_p^2}\:\varrho,\qquad 3H^2+2\dot H={}-\frac{1}{m_p^2}\:p.
\label{eomprho}
\end{equation}
Here $H$ is the Hubble parameter: $H\equiv \dot a(t)/a(t)$ and the dot denotes the
time derivative. If the scalar field depends only on time, i.e.
$\phi=\phi(t)$, then
\begin{equation}
\varrho={}-\frac12\dot\phi^2+V(\phi),\qquad p={}-\frac12\dot\phi^2-V(\phi).
\label{prho}
\end{equation}
We recast system (\ref{eomprho}) to the following form
\begin{equation}
\dot H=\frac{1}{2m_p^2}\dot\phi^2,\qquad
3H^2=\frac{1}{m_p^2}\left(V(\phi)-\frac{1}{2}\dot\phi^2\right).\label{eom12}
\end{equation}
Besides of equations (\ref{eom12})  one can obtain from the action (\ref{action})
an equation of motion for the field $\phi$
\begin{equation}
\ddot{\phi}+3H\dot\phi-V_{\phi}^{\prime}=0, \label{eomphi2order}
\end{equation}
where $V_{\phi}^{\prime}\equiv\frac{dV}{d\phi}$. This equation is in fact a
consequence of system (\ref{eomprho}). Expressing $H$ through $\phi$ and
$\dot\phi$, we obtain
\begin{equation}
\ddot{\phi}+\sign(H)\frac{\sqrt{3}}{m_p}\dot\phi
\sqrt{-\frac12\dot\phi^2+V(\phi)}-V_{\phi}^{\prime}=0.
\label{eomphi}
\end{equation}
Here $\sign(H)$ factor selects the proper branch of the square root.  The
expanding Universe corresponds to $H>0$.

\subsection{Connection with the String Theory}

At present time one of the possible scenarios of the Universe evolution considers
the Universe to be a D3-brane (3 spatial and one time variable) embedded in
higher-dimensional space-time. D-branes do naturally emerge in the open string
theory (in our case we consider fermionic open strings containing both GSO$+$
and GSO$-$ sectors). The D-brane in question is unstable and does evolve to
the stable state. This process is described by the open string dynamics,
which ends are attached to the brane (see reviews \cite{review-sft}  and
references therein). If only the lowest excitation
--- the tachyon --- is taken into account then the D-brane dynamics are described
by an action of the open string tachyon. There are two common ways to describe
the tachyon behavior: Dirac-Born-Infeld (DBI) approximation and the level
truncation scheme in the covariant string field theory \cite{Witten,AMZ,PTY}. The
level truncation method advocated itself checking the Sen's conjecture and,
therefore, it is natural to use it to analyze the dynamics of the D-brane, see
\cite{Zw,AJK,yar}. The action for the tachyon resulting from the fermionic string
field theory in the approximation of the slow-varying auxiliary
field~\cite{AJK} has the form
\begin{equation*}
S_{\text{flat}}=\frac1{g_o^2}\int dx\left(
-\frac12\partial_{\mu}\varphi\partial^{\mu}\varphi
+\frac14\varphi^2-\frac{81}{256}\tilde{\varphi}^4\right).
\end{equation*}
where $\varphi$ is the dimensionless tachyon field and the coordinates
and the constant $g_o$ are also dimensionless. The tilde means an action
of the non-local operator
$\exp\left(-\log\left(\frac4{3\sqrt{3}}\right)\partial^2\right)$ on the
field $\varphi$.
The subscript ``flat'' hereafter indicates the flat background
metric is used. Note that in the bosonic string case the interaction
is cubic. For the space homogenous configurations only the dependence
on time $\tau$ is left and the action takes the form
\begin{equation*}
S_{\text{flat}}^{(\tau)}=\frac1{g_o^2}\int d\tau\left(
+\frac12\left(\frac{d\varphi}{d\tau}\right)^2
+\frac14\varphi^2-\frac{81}{256}\left(e^{\log\left(\frac4{3\sqrt{3}}\right)\frac{d^2}{d\tau^2}
}\varphi\right)^4\right).
\end{equation*}
Introducing the notation
$\phi=\frac9{4\sqrt{2}}\exp\left(\log\left(\frac4{3\sqrt{3}}\right)\frac{d^2}{d\tau^2}\right)\varphi$
and  $t=\tau/2$, we rewrite  the
equations of motion obtained from the last action in the following form
\begin{equation*}
\left(-\frac{d^2}{dt^2}+2\right)e^{\tilde{a}\frac{d^2}{dt^2}}\phi=2\phi^3,
\end{equation*}
where $\tilde{a}=-\frac14\log\left(\frac4{3\sqrt{3}}\right)>0$. In fact this
equation is integral one since the exponent containing an infinite number of
derivatives can be replaced with an integral operator \cite{Zw} in some class
of fields as follows
\begin{equation}
\left(-\frac{d^2}{dt^2}+2\right)\frac1{\sqrt{4\pi \tilde{a}}}\int
e^{\frac{\left(t-t^{\prime}\right)^2}{4\tilde{a}}}f(t^{\prime})dt^{\prime}=2\phi(t)^3.
\label{flattinteom}
\end{equation}
Note, that equation (\ref{flattinteom}) without the first summand is an
equation for the $p$-adic string (in our case $p=3$)~\cite{padic}. In the
latter case the existence of a solution interpolating between vacua has been proved
\cite{VV}. Previously this solution has been numerically constructed
in~\cite{BFOW}.

Equation (\ref{flattinteom}) was investigated in \cite{AJK,yar} and an
existence of a rolling solution, i.e. a solution interpolating between the two
vacua of the potential
\begin{equation}
V_{\text{flat}}(\phi)=\frac12\left(1-\phi^2\right)^2 \label{flatV}
\end{equation}
was shown numerically. It is remarkable that the late time behavior of the rolling
solution can be effectively described by a lagrangian with a ghost sign of the
kinetic term and the same potential. The corresponding action has the form
\begin{equation}
S_{\text{eff,flat}}^{(t)}= \int dt \left(
 -\frac{1}{2}\dot\phi^2
-\frac12\left(1-\phi^2\right)^2\right).
\label{flatteffact}
\end{equation}
As known, the equation of motion derived from the action
(\ref{flatteffact}):
\begin{equation}
\ddot\phi+2\phi\left(1-\phi^2\right)=0 \label{flateom}
\end{equation}
has an interpolating solution --- a kink
\begin{equation}
\phi(t)=\tanh(t). \label{kinkphi}
\end{equation}

\subsection{Construction of solution}

System of equations (\ref{eom12}) with some polynomial potential $V(\phi)$ is
not integrable. The numeric analysis of the Friedmann equations with the potential
(\ref{flatV}) has been performed in the paper~\cite{Arefeva}. It is known (
see, for example~\cite{Padmanabhan,DeWolfe}) that it is always possible to
find the potential $V(\phi)$, if either $\phi(t)$ or $H(t)$ are given explicitly.

It is worth to note that our model has common features with models considered
in brane-world scenarios (see Appendix). Following~\cite{DeWolfe}, we assume,
that $H(t)$ is a function of $\phi(t)$, called superpotential, that is
$H(t)=W(\phi(t))$.  Using equality $\dot H=W_{\phi}^{\prime}\dot\phi$, we
obtain from system (\ref{eom12}):
\begin{eqnarray}
&&\dot\phi=2m_p^2 W_\phi^{\prime},\label{eom1W}
\\&&V=2m_p^4{W_\phi^{\prime}}^2+3m_p^2W^2. \label{eom2W}
\end{eqnarray}

Let us find a potential, which corresponds to the field $\phi$ of
type (\ref{kinkphi}). This function satisfies the following equation
\begin{equation}
\dot\phi=1-\phi^2, \label{dotphi}
\end{equation}
hence, from (\ref{eom1W}) it follows
\begin{equation*}
W=\frac{1}{2m_p^2}\left(\phi-\frac{1}{3}{\phi}^{3} \right)+ C,
\end{equation*}
where $C$ is an integration constant. Once $W(\phi)$ is known we obtain the
potential $V(\phi)$ from (\ref{eom2W}). Different values of $C$ correspond to
different $V(\phi)$. The requirement that $V(\phi)$ is an even function is
equivalent to the condition $C=0$. In this case
\begin{equation}
V(\phi)= \frac{1}{2}\left(1-\phi^2\right)^2+
\frac{1}{12m_p^2}\phi^2\left(3-\phi^2\right)^2. \label{ourV}
\end{equation}

It can be straightforwardly verified that if $\phi(t)$ is a solution
to (\ref{eom1W}), (\ref{eom2W}) with some potential $V(\phi)$, then $\phi(-t)$ is a
solution as well. For example, $\phi_2(t)=\tanh(-t)=-\tanh(t)$ also
is a solution to (\ref{eom1W}), (\ref{eom2W}) with the potential (\ref{ourV}). Indeed
these two solutions correspond to the one and the same function
$H(t)$, but to the two different functions $W(\phi)$.

We have constructed the potential $V$, using an explicit form of a solution.
Now we consider more general problem and look for solutions, starting with the
the following two requirements: $\phi(t)$ is a smooth function with nonzero
asymptotics $\phi(\pm\infty)=\pm B$ and the superpotential $W(\phi)$ is a
polynomial in $\phi$ not more than the third degree. In the first condition we
can put $B=1$ without loss of generality. The second condition guarantees that
the potential $V(\phi)$ will be a polynomial in $\phi$ and its  degree is not
more than six. The above conditions allow us to rewrite equation (\ref{eom1W})
in the following form:
\begin{equation}
\dot\phi=\alpha+\beta\phi+\gamma\phi^2.
\label{subs}
\end{equation}
From the asymptotic conditions it follows
\begin{equation}
\gamma=-\alpha,\qquad\beta=0.
\label{conditions}
\end{equation}
Taking into account (\ref{conditions}) we solve equation (\ref{subs}) and
obtain
\begin{equation}
\phi(t)=\tanh(\alpha (t-t_0)). \label{tanh}
\end{equation}
We do not consider another solution: $\phi(t)=\coth(\alpha (t-t_0))$, because
it is not a smooth function. Thus, up to a scaling and a time shift we obtain a
solution, which coincides with the solution in the flat metric. It is evident
that the potential obtained from (\ref{eom2W}) coincides with
(\ref{ourV}) up to an overall factor.  It is interesting that under
above-mentioned conditions on a solution and a superpotential, the sixth degree
potential is a minimal possibility for existence of an interpolating solution.
One can also see, that any solution to the equations of motion in flat space with
the fourth degree potential is a solution to the Friedmann equations either
with the sixth degree polynomial potential or with a nonpolynomial
potential~\cite{vernov05}.

It is interesting to note that we have built an exact solution without any
approximation. One of the standard approaches to the analysis of cosmological
solutions in the presence of a scalar field is the so called slow-roll
approximation which technically means a neglecting of the terms $\ddot\phi$ in
equation (\ref{eomphi2order}) and $\dot\phi^2$ in the expression for the
energy (\ref{prho}). However,  from (\ref{dotphi}) it follows that for our
solution the following relations take place:
\begin{equation*}
\ddot\phi =-2\phi \left(1-\phi^2\right),\qquad
\dot\phi^2=\left(1-\phi^2\right)^2.
\end{equation*}

It is evident that the slow-roll approximation in this case is valid only when
$\phi$ is close to~$1$. The above-mentioned approximation describes our
solution at very large times and can not be used for its description in the
beginning (close to zero) moments of time.

The Hubble parameter for the constructed solution is given by
\begin{equation*}
H=\frac{1}{2m_p^2}\phi\left(1-\frac{1}{3}{\phi}^{2}
\right)=\frac{1}{2m_p^2}\tanh(t)\left(1-\frac{1}{3}\tanh(t)^2\right).
\end{equation*}
This parameter goes asymptotically to $1/(3m_p^2)$ when
$t\rightarrow\infty$, or equivalently $\phi\rightarrow 1$. Once
$H(t)$ is known one readily obtains the scale factor
\begin{equation}
a=a_0\frac{e^{\phi^2/(12m_p^2)}}{(1-\phi^2)^{1/(6m_p^2)}}
=a_0(\cosh(t))^{\frac{1}{3m_p^2}}\exp\left({\frac{\cosh(t)^2-1}
{12m_p^2\cosh(t)^2}}\right), \label{func_at}
\end{equation}
where $a_0$ is an arbitrary constant.


\section{COSMOLOGICAL CONSEQUENCES}

\subsection{Acceleration}

 The function $a(t)$ has the following asymptotic behavior at large $t$:
\begin{equation}
\lim_{t\to\infty}a(t)\sim e^{\frac{1}{3m_p^2}t}.
\label{func_a_limit}
\end{equation}
It follows from formulae (\ref{func_at}) and (\ref{func_a_limit})
that the Universe expands with an acceleration. The deceleration
parameter is negative and equal to
\begin{equation*}
q(t)=-\:\frac{\ddot{a}a}{\dot{a}^2}=-\:\frac2{\sinh(t)^2}
-\frac{5+\cosh(4t)}{2(2+\cosh(2t))^2}.
\end{equation*}

\begin{figure}[h]
\centering
\includegraphics[width=50mm]{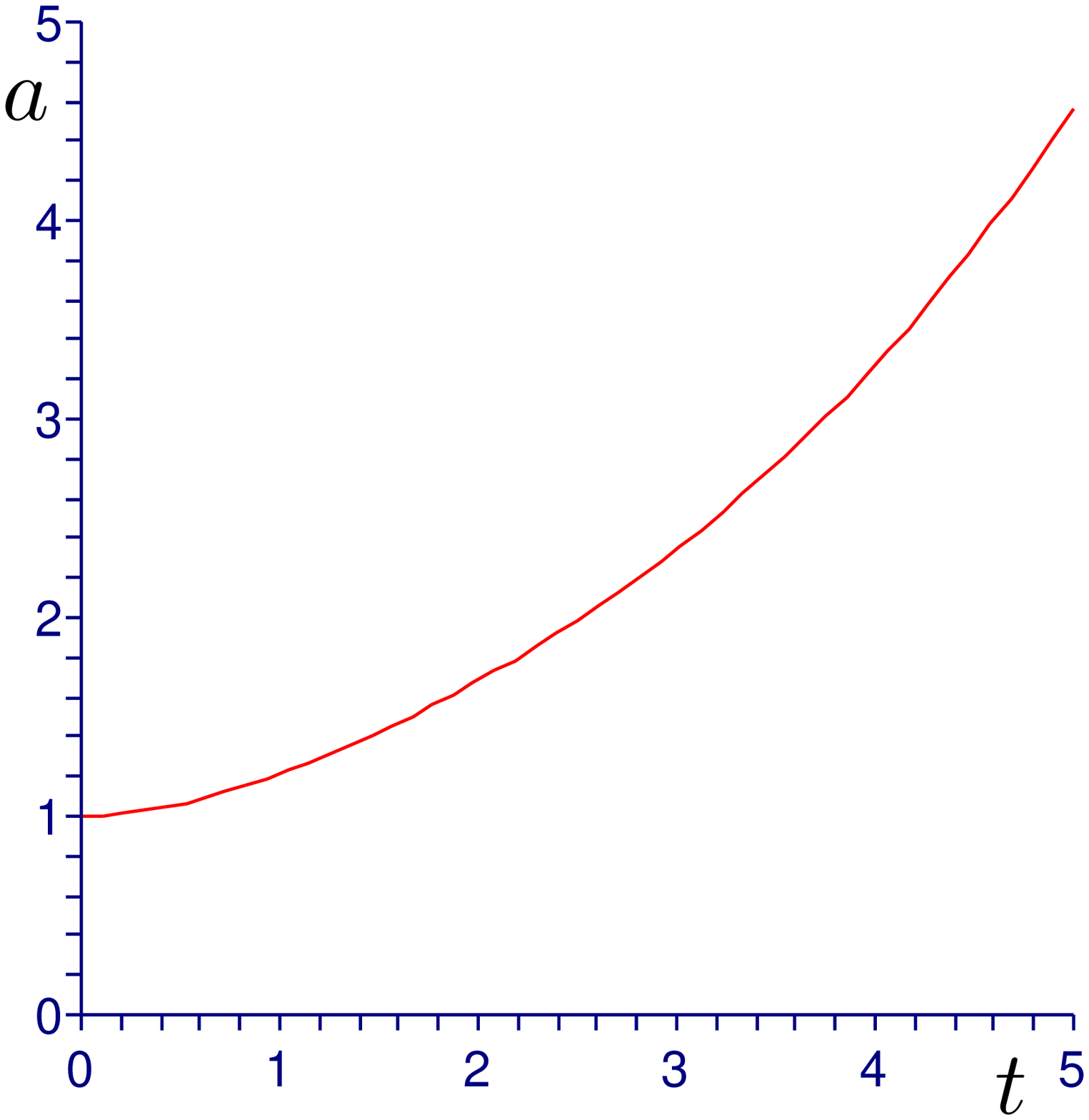} { \ \ \ }
\includegraphics[width=50mm]{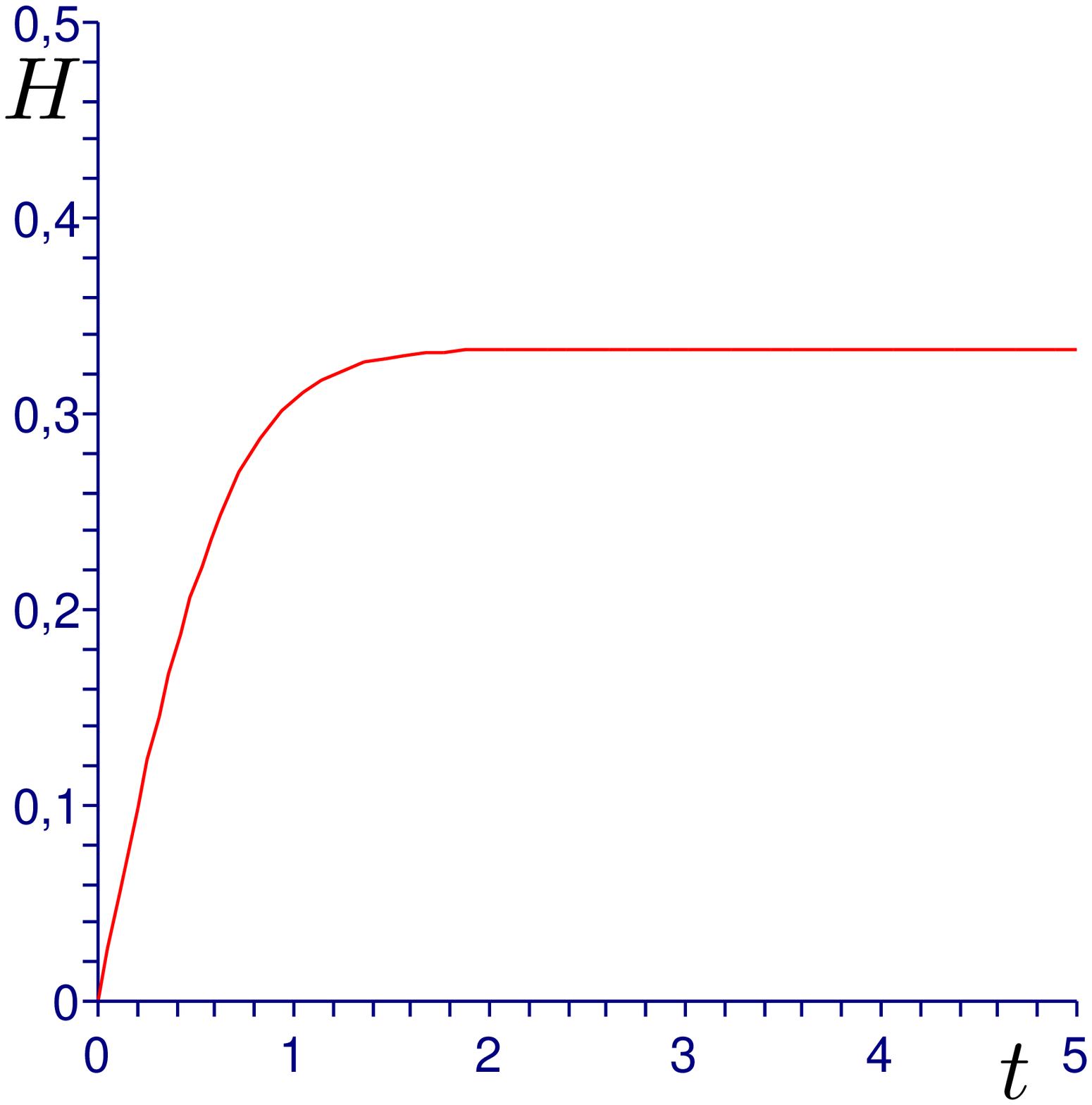} { \ \ \ }
\includegraphics[width=50mm]{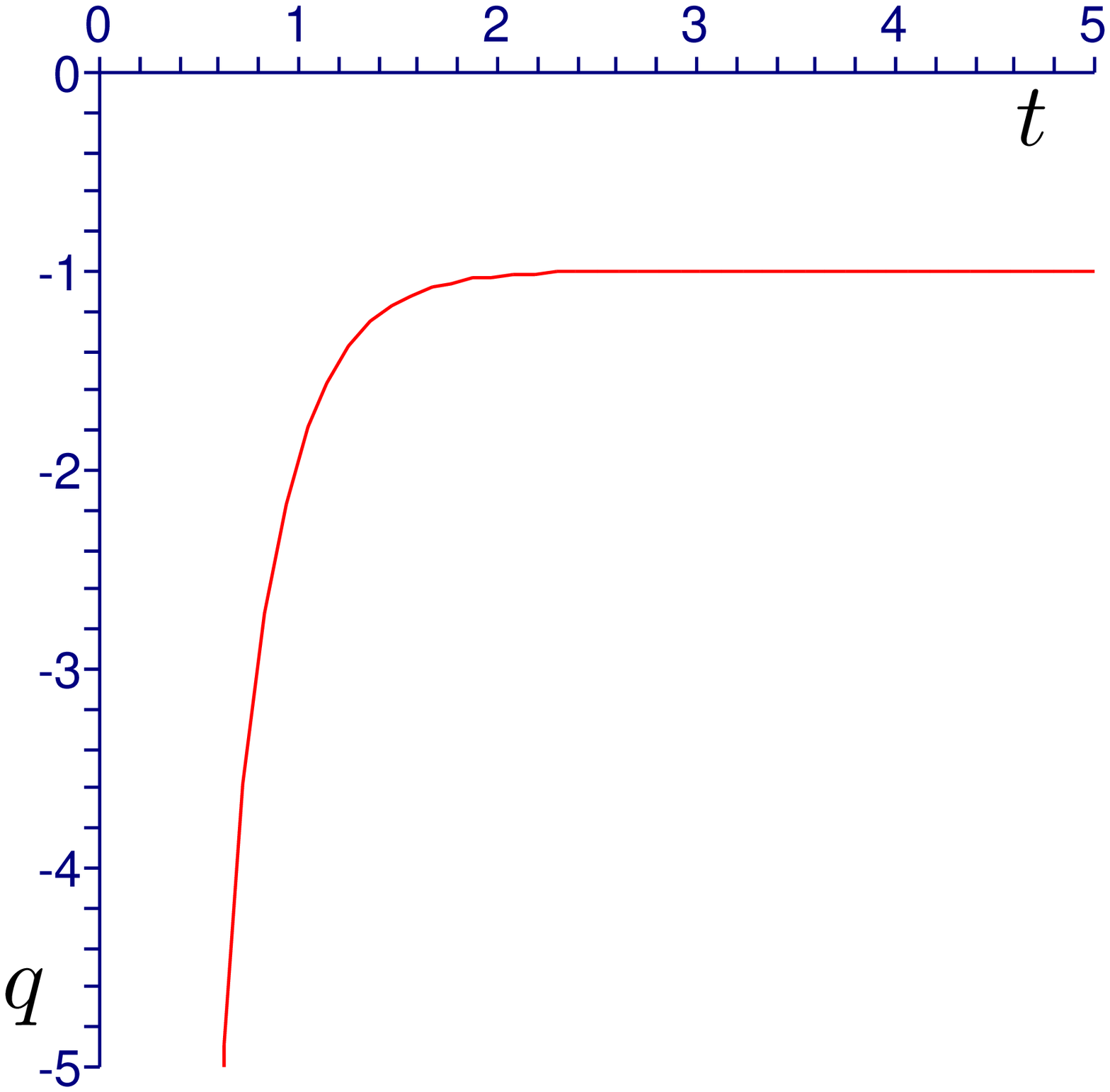}\\[1mm]
\caption{The scale factor $a(t)$ (left), $a_0=1$, the Hubble parameter $H(t)$
(center) and the deceleration parameter $q$ (right)}  \label{Ha}
\end{figure}

\begin{figure}[h]
\centering
\includegraphics[width=50mm]{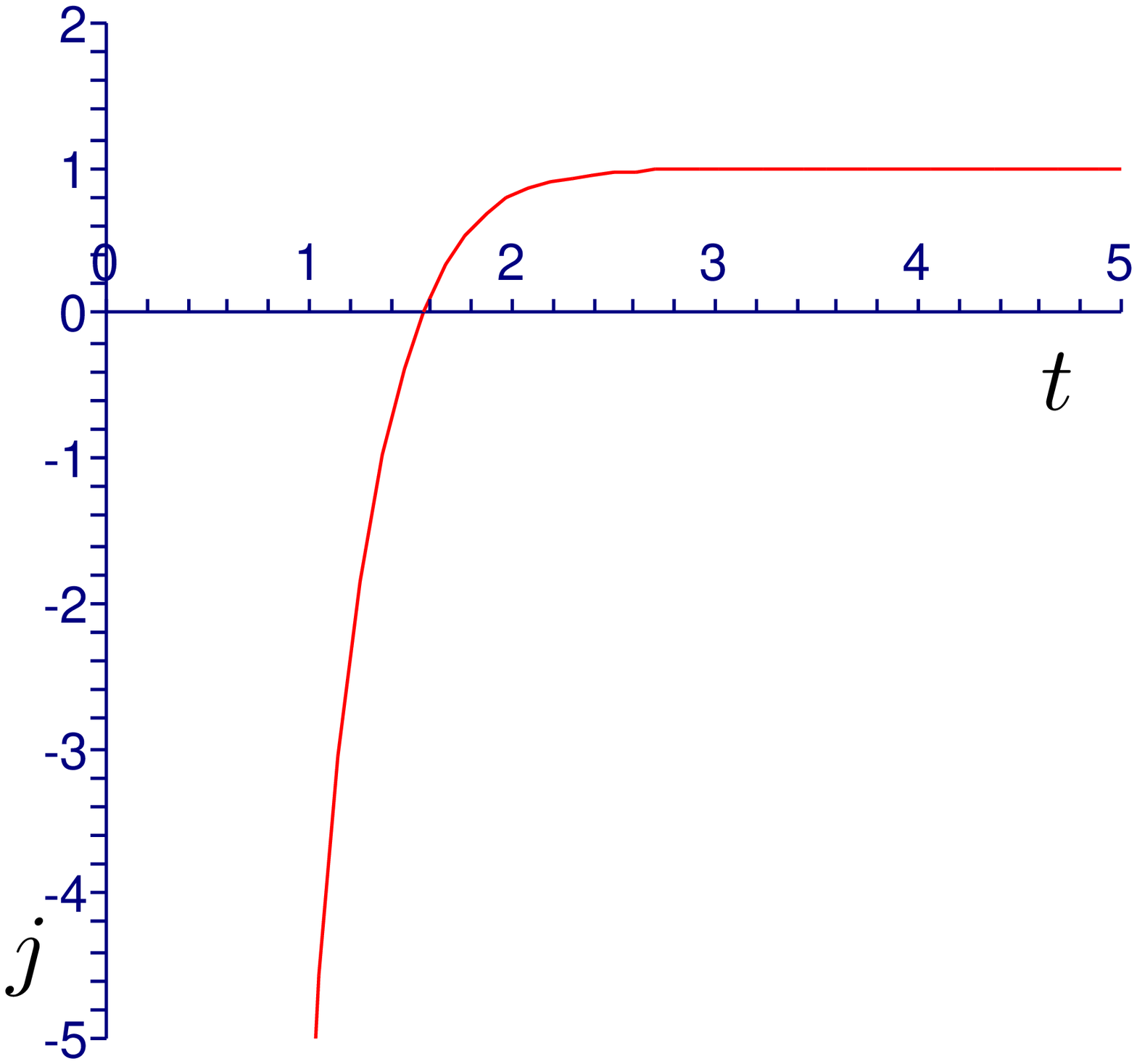} { \ \ \ }
\includegraphics[width=50mm]{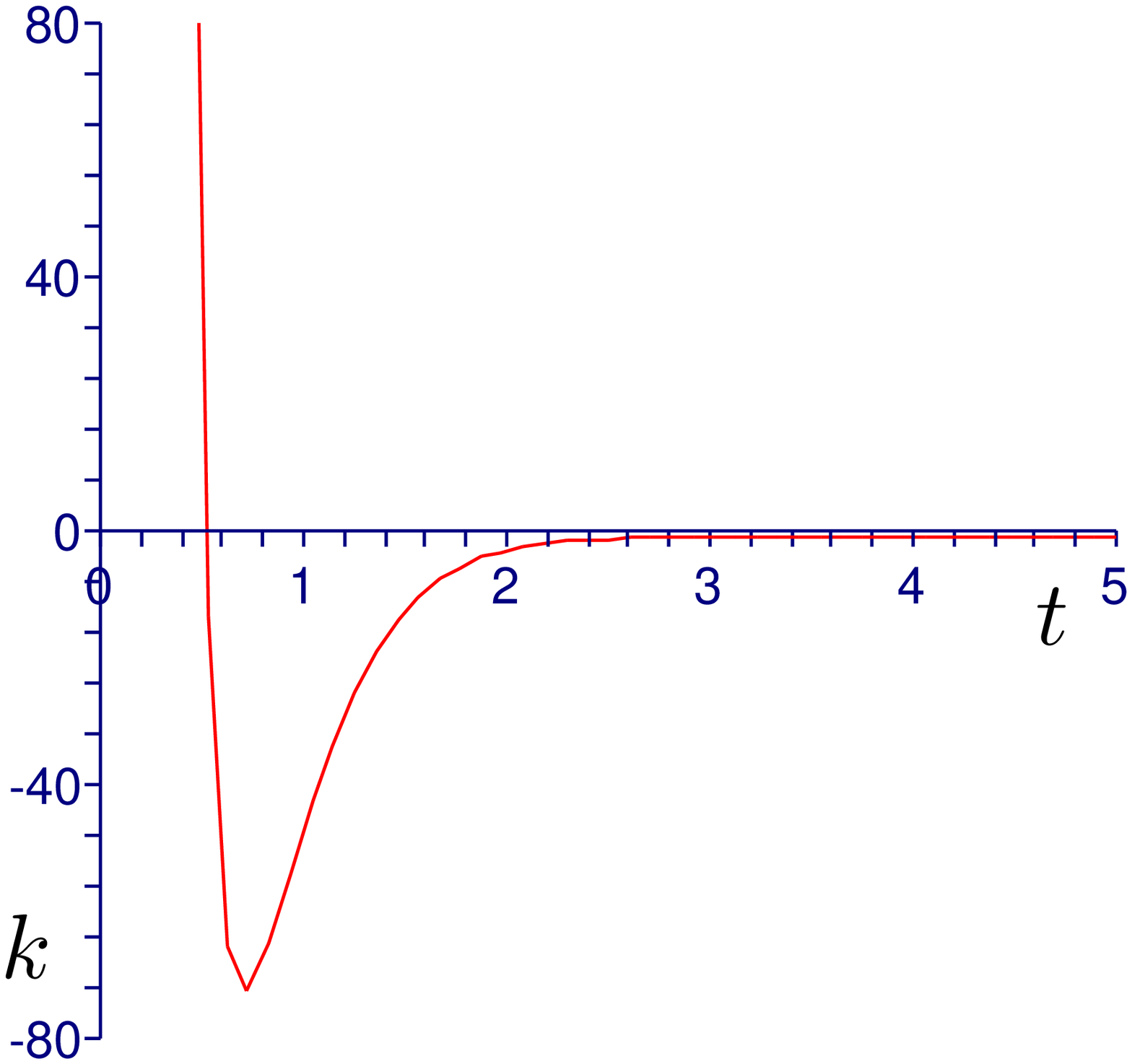} { \ \ \ }
\includegraphics[width=50mm]{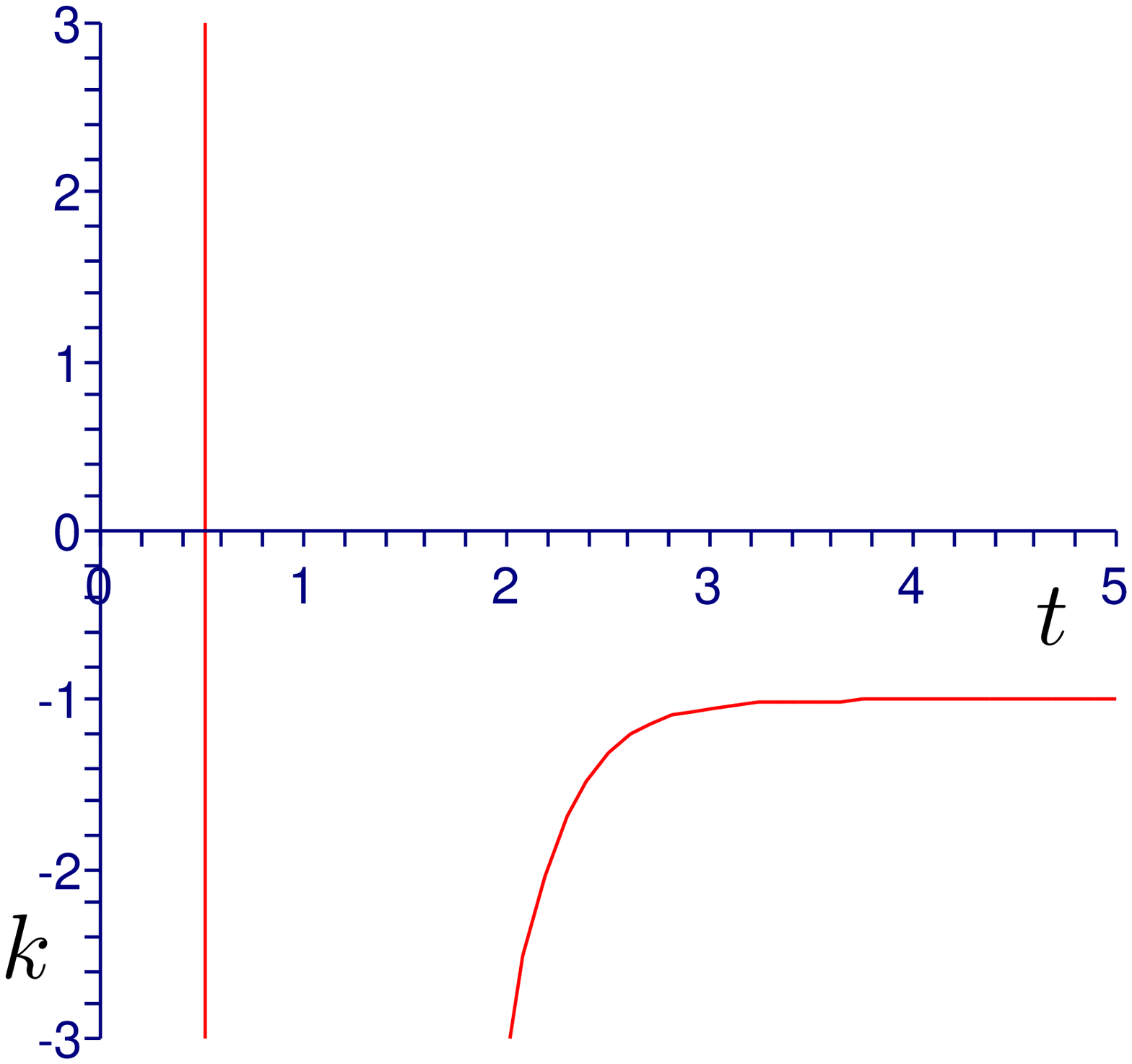}\\[1mm]
\caption{``jerk'' parameter (left), ``kerk'' parameter (center) and its fine
structure (right).}  \label{qjk}
\end{figure}

The corresponding plots are drawn in Fig. \ref{Ha} (Hereafter we assume
$m_p=1$ for all plots if something else is not specified explicitly).
 The ``jerk'' parameter
$j=\dot{\ddot{a}}a^2/\dot{a}^3$ and the ``kerk'' parameter
$k=-\ddot{\ddot{a}}a^3/\dot{a}^4$ are presented in Fig.~\ref{qjk}.

Substituting the obtained solution (\ref{kinkphi}) and the potential
(\ref{ourV}) into expressions for the pressure and energy densities, we obtain
\begin{equation*}
p(\phi)=-\left(1-\phi^2\right)^2-\frac1{12m_p^2}
\phi^2\left(3-\phi^2\right)^2,\qquad
\varrho(\phi)=\frac1{12m_p^2}\phi^2\left(3-\phi^2\right)^2.
\end{equation*}
Therefore, the parameter of state $w$ is given by
\begin{equation*}
w\equiv\frac {p}{\varrho}=-1-12m_p^2\frac{\left(1-\phi^2\right)^2}{\phi^2
\left(3-\phi^2\right)^2}=-1-12m_p^2\frac{\cosh(t)^2}{\sinh(t)^2\left(1+2\cosh(t)^2
\right)^2}.
\end{equation*}

\begin{figure}[h]
\centering
\includegraphics[width=50mm]{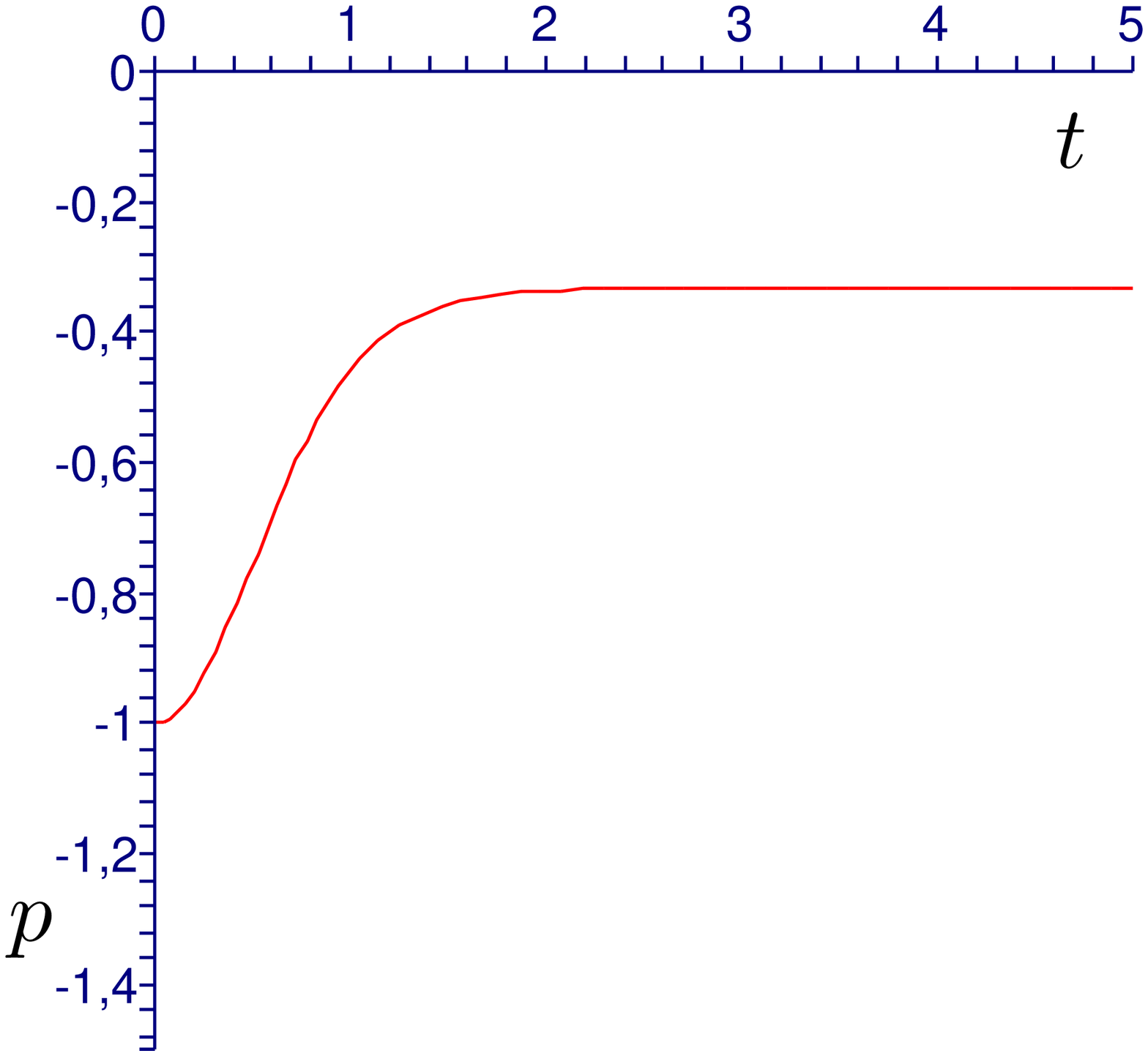} { \ \ \ }
\includegraphics[width=50mm]{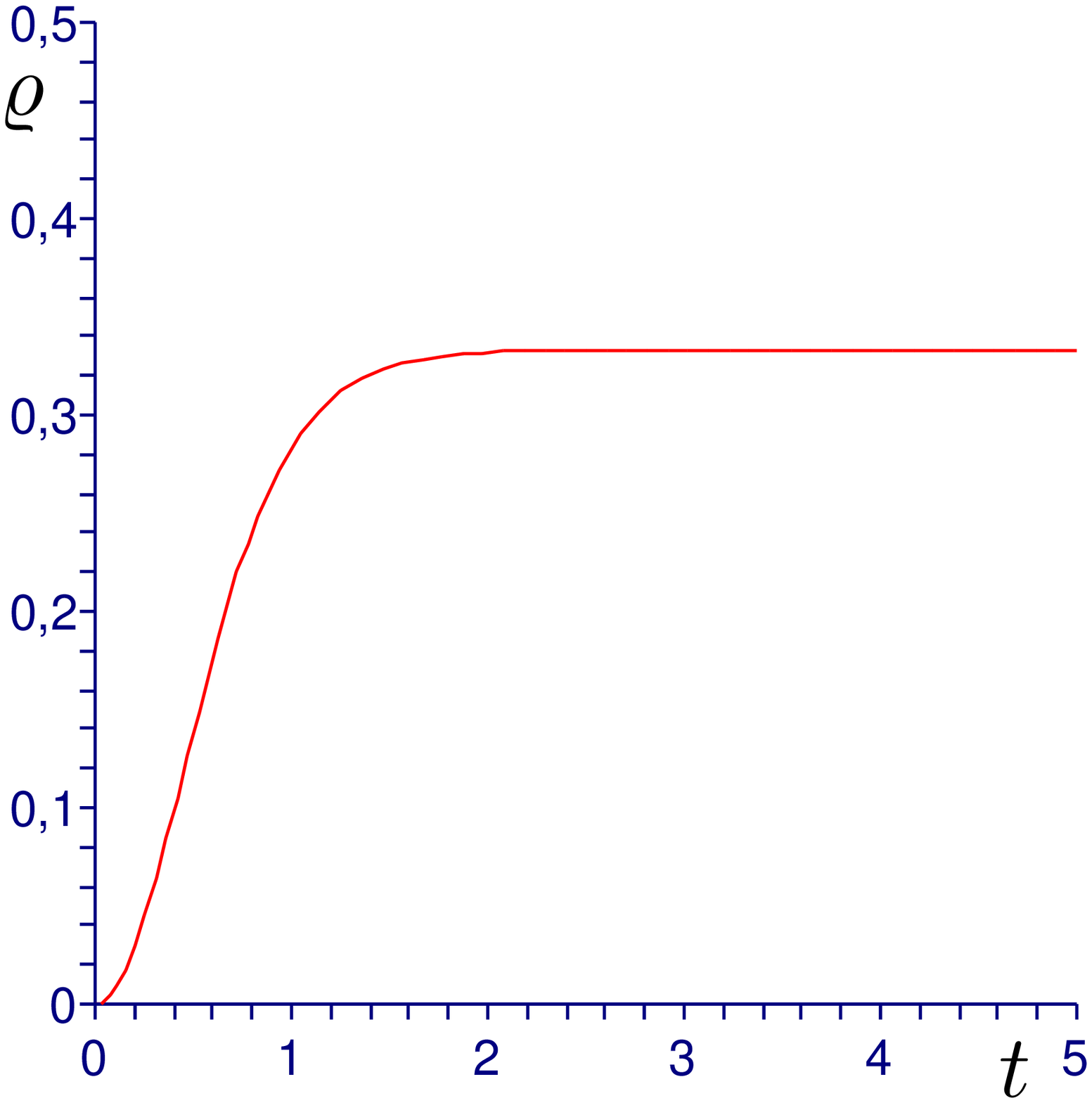} { \ \ \ }
\includegraphics[width=50mm]{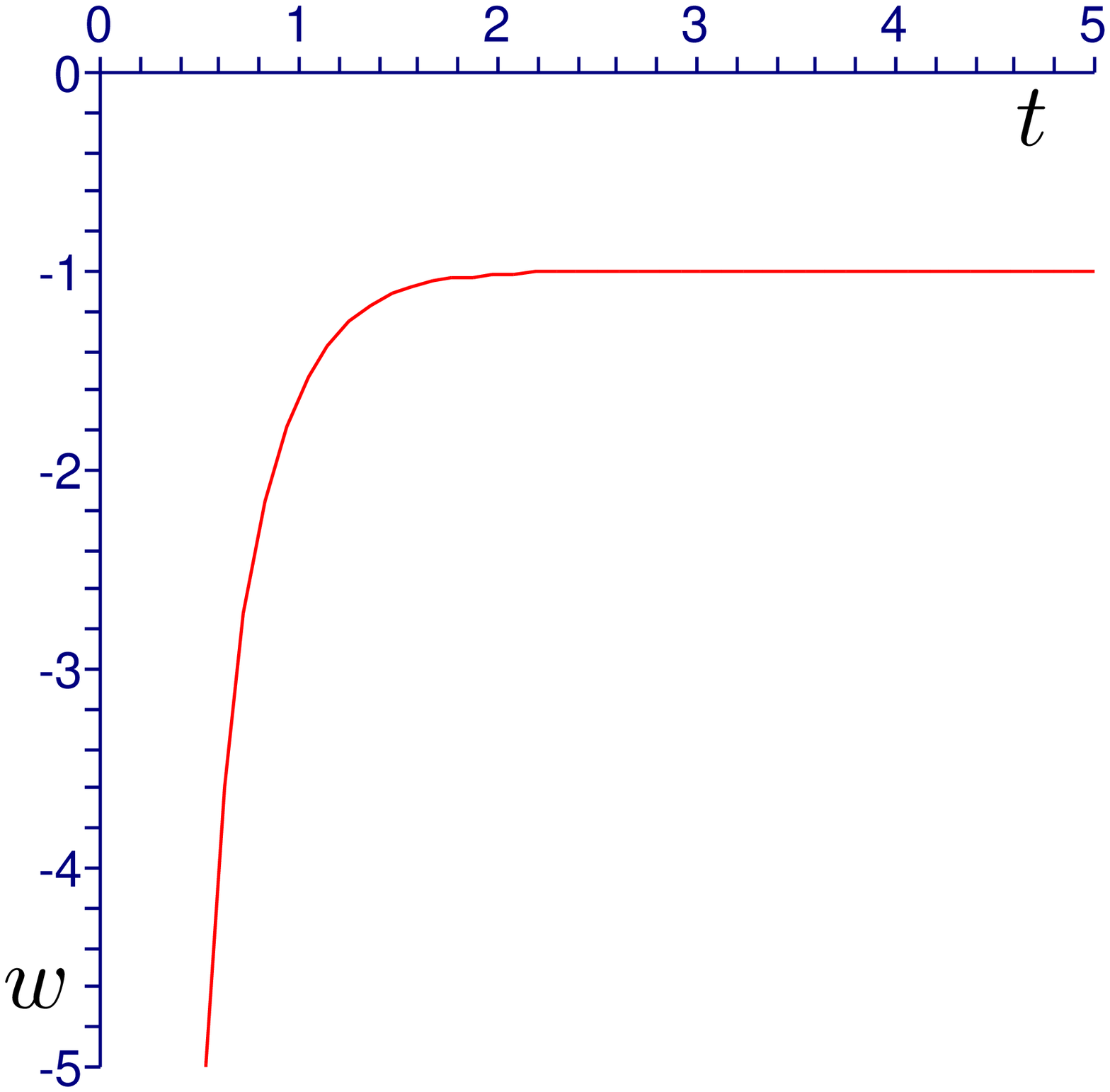}
\caption{$p(t)$ (left), $\varrho(t)$ (center) and $w(t)$ (right)}
\label{prho_fig}
\end{figure}

Note that the function $w(\phi)$ is less than $-1$ for $|\phi|\neq 1$ and is
equal to $-1$ for $|\phi|=1$. Point $\phi=1$ corresponds to infinite future.
The plots for the pressure density $p(t)$, the energy density $\varrho(t)$  and
the parameter of state $w$ are presented in Fig. \ref{prho_fig}.

Thus, in the considered model of the accelerated expanding Universe the
phantom field describes the dark energy with the state parameter $w<-1$.


\subsection{Evolution of the exact solution to equation (\ref{eomphi})}

Solution (\ref{kinkphi}) to equation (\ref{eomphi}) is a function describing
an interpolation of the field $\phi$ from point $-1$ to point $1$ (and
\textit{not} viceversa) with zero initial and final velocities in the
potential (\ref{ourV}) with a friction proportional to
$\sign(H)\sqrt{-\frac12\dot\phi^2+V(\phi)}$, which depends on coordinate and
velocity. Factor $\sign(H)$ makes the friction negative for negative $\phi$,
i.e. the particle is accelerated by it. Indeed, the expression for friction
coefficient is equal to $\phi\left(3-\phi^2\right)/(2\sqrt{3}m_p)$ on the
solution for both positive and negative $\phi$.

\begin{figure}[h]
\centering
\includegraphics[width=72mm]{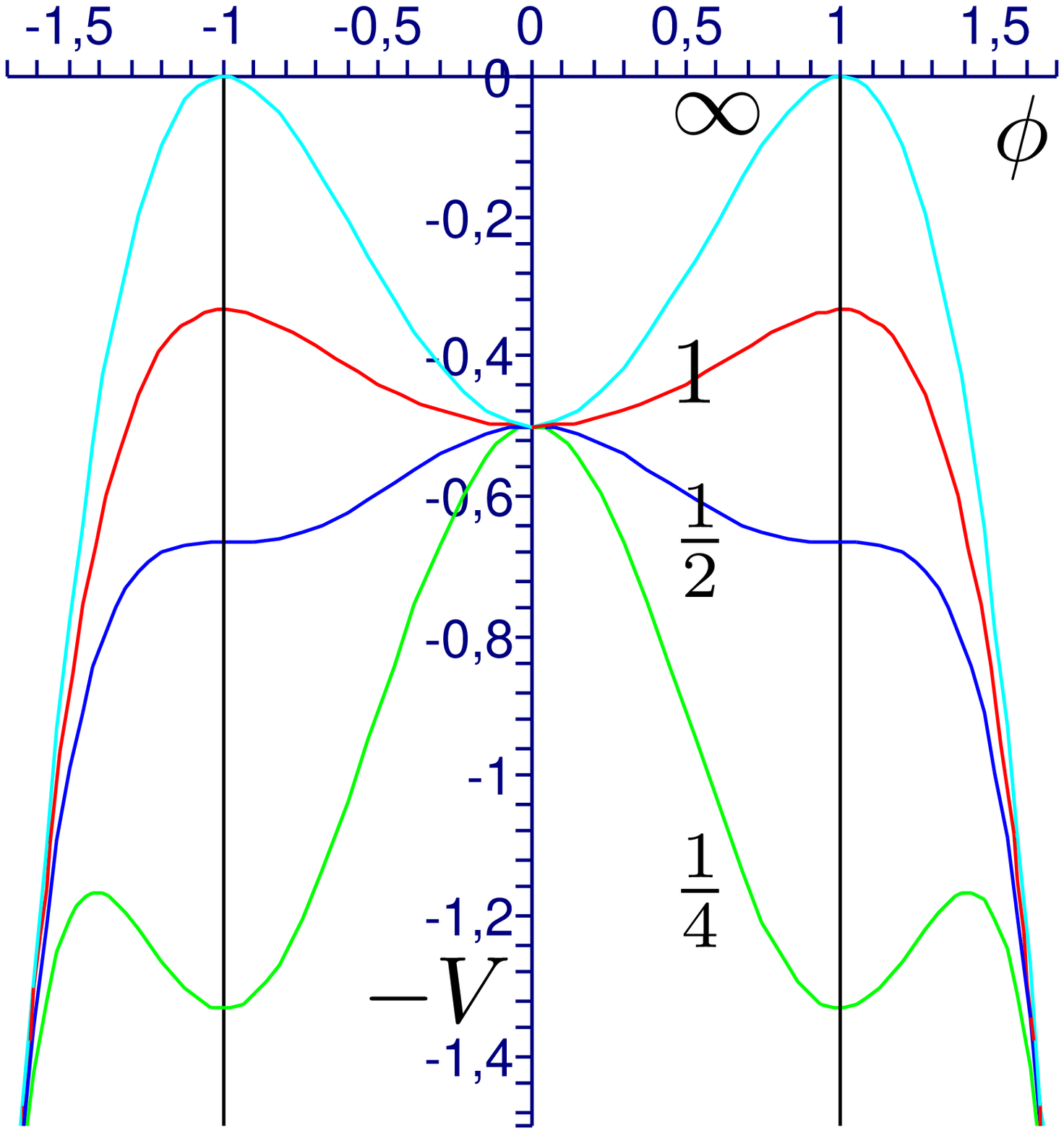}{ \ \ \ \ }
\includegraphics[width=72mm]{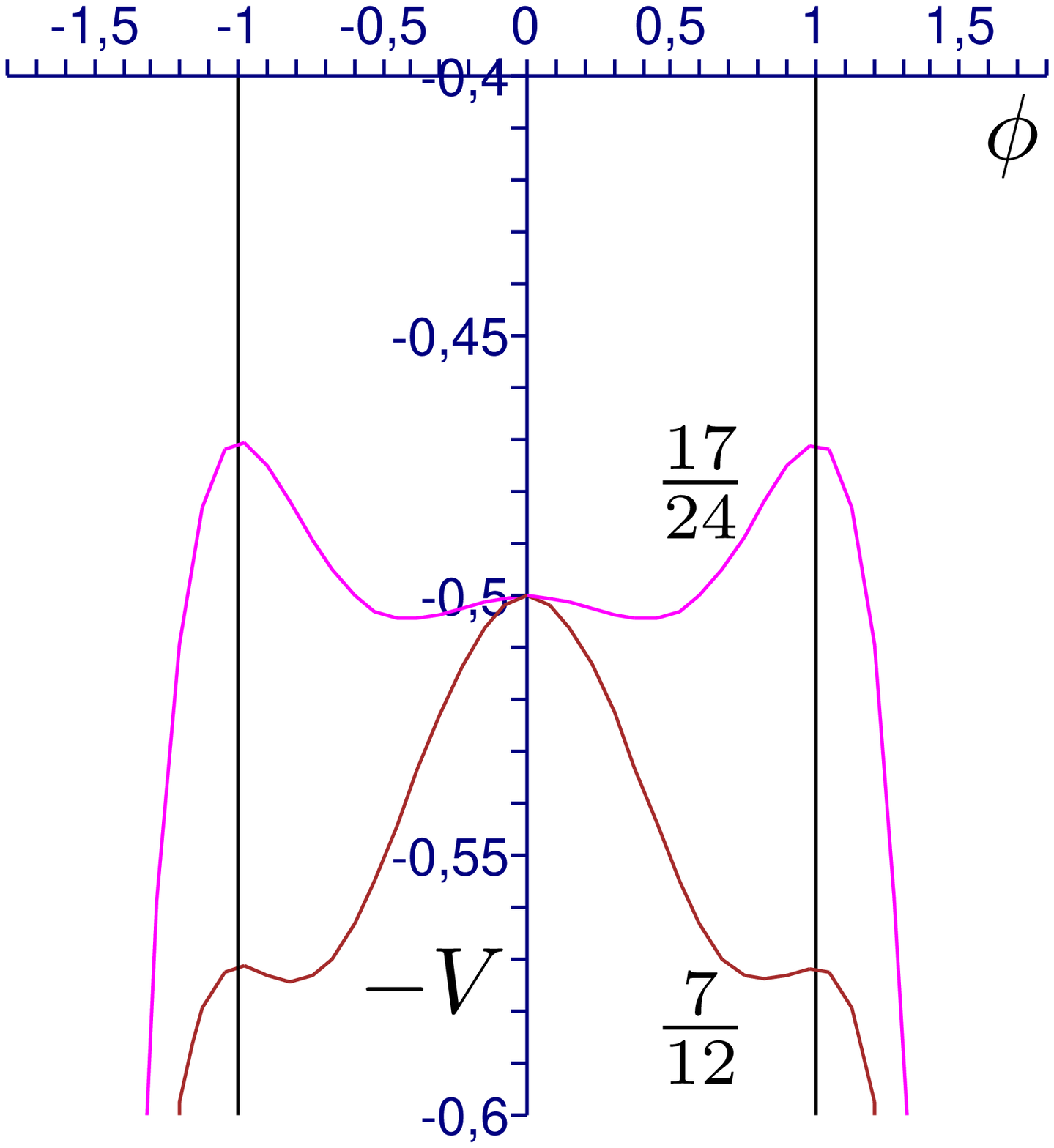}
\caption{Potential $-V$ for different values of $m_p$. The following $m_p$
correspond to the curves (top-down direction): $m_p^2=\infty$,
 $m_p^2=1$, $m_p^2=1/2$ and $m_p^2=1/4$ in the left plot and
 $m_p^2=17/24$ and $m_p^2=7/12$ in the right plot.
Two vertical lines $\phi=\pm 1$ are used as a grid.} \label{Vflipplot}
\end{figure}

Let us discuss on the evolution of the exactly constructed solution.
A phantom field evolution is equivalent to an evolution of the
ordinary field in the flipped potential. The shape of potential $V$
essentially depends on value of $m_p$, see Fig.~\ref{Vflipplot},
where the flipped potential $-V$ is plotted for different values of
$m_p$.

To analyze the extrema, let us note the following expressions
\begin{equation}
\begin{split}
V(0)&=\frac12,\qquad V(\pm 1)=\frac1{3m_p^2},\qquad
V^{\prime}_{\phi}=\frac1{2m_p^2}\phi\left(1-\phi^2\right)\left((3-4m_p^2)-\phi^2\right),\\
V^{\prime\prime}_{\phi}(0)&=\frac{3-4m_p^2}{2m_p^2},\qquad
V^{\prime\prime}_{\phi}(\pm 1)=\frac{2\left(2m_p^2-1\right)}{m_p^2}.
\end{split}
\label{Vdetails}
\end{equation}
It is evident from (\ref{Vdetails}) that extrema $\phi=\pm 1$ and $\phi=0$ are
independent of $m_p$, whereas other extrema are functions of $m_p$. Moreover,
a particular value of $m_p$ determines whether a specific extremum is a
maximum or a minimum. Possible values of $m_p$ can be separated in the
following domains which determine a structure of the flipped potential $-V$
extrema.

\begin{itemize}

\item For $m_p=\infty$ $V(\phi)$ is a well known double well
potential.

\item For $\frac34\leqslant m_p^2<\infty$ the flipped potential $-V$,
being a 6th degree potential is similar graphically to a flipped
double well potential: it has a local minimum in the point $\phi=0$
and maxima in the points $\phi=\pm1$.

\item For $\frac12 < m_p^2<\frac34$ point
$\phi=0$ in the flipped potential becomes a local maximum two more
minima appear in the interval $(-1,1)$, developing two wells and one
hill on the way of our field $\phi$ during an interpolation from
point $-1$ to point $1$, see the right plot in Fig.~\ref{Vflipplot}.
In more details:

\begin{itemize}

\item For $\frac23\leqslant m_p^2<\frac34$ points $\phi=\pm 1$ are not
below the point $\phi=0$ in the flipped potential.

\item For $\frac12< m_p^2<\frac23$ points $\phi=\pm 1$ are below the
point $\phi=0$ in the flipped potential. This means that our field
$\phi$ starts at the point $\phi=-1$ with zero initial velocity,
rolls down to the well and then climbs up to the hill which is above
its initial location. This is counterintuitive while the factor
$\sign(H)$ in the friction coefficient is not taken into account.
The latter means the friction is negative for negative $\phi$. It
seems one can disprove the last statement since it is natural to
inverse the time and consider the backward motion. It is easy to
resolve this contradiction because in the case of an inverse time
function $H$ would have an opposite sign.

\end{itemize}

\item For $m_p^2=\frac12$ the flipped potential has the only extremum
which is maximum in the origin and points $\phi=\pm1$ are inflection
points.

\item For $m_p^2<\frac12$ two minima, coming from the last multiplier of the
expression $V^{\prime}$ (\ref{Vdetails}), go beyond the interval $-1<\phi<1$
and become maxima. Points $\phi=\pm 1$ become minima of the flipped potential.
Point $\phi=0$ remains the maximum. In this case field $\phi$ starting at
point $\phi=-1$ climbs up to the hill from the very beginning and then rolls
down and stops at the point $\phi=1$. Such a behavior seems incredible and
becomes real due to a negative friction for negative $\phi$.

\end{itemize}

Investigating the cosmological evolution of the phantom field we use only the
a part of solution (\ref{kinkphi}) starting at the time $t=t_{in}>0$. At this
time the field is at the point $\phi_{in}=\phi(t_{in})=\tanh (t_{in})$ and one
has to supply an initial velocity
$\dot\phi(t_{in})=1-\phi_{in}^2=1/\cosh^2(t_{in})$. For large $m_p^2$ (weak
coupling to the gravity) the phantom field uses an initial kinetic energy to
climb up to the hill at $\phi=1$. Increasing a coupling to gravity, i.e.
decreasing $m_p^2$ we lower the height of the hill at the point $\phi=1$ of
the flipped potential. For $m_p^2=2/3$ heights of the hills at points $\phi=0$
and $\phi=1$ become equal and all initial kinetic energy is spent for the work
against the friction. For $m_p^2<2/3$ the friction becomes stronger and the
particle has the kinetic energy which is exactly enough to reach the point
$\phi=1$ which is now lower than the hill at point $\phi=0$.

\section{STABILITY OF SOLUTIONS}
\subsection{Variation of initial data}
Let us consider the behavior of the solution to system (\ref{eom12}) in the
neighborhood of our exact solution
\begin{equation*}
\phi_0(t)=\tanh(t),\qquad
H_0(t)=\frac{1}{2m_p^2}\tanh(t)\left(1-\frac{1}{3}\tanh(t)^2\right),
\end{equation*}
with the aim to analyze its stability. Substituting
$\phi(t)=\phi_0(t)+\varepsilon\phi_1(t)$ and $ H(t)=H_0(t)+\varepsilon H_1(t)$
in system (\ref{eom12}), we obtain in the first order of $\varepsilon$ the
following equations:
\begin{equation}
\begin{array}{@{}rcl@{}}
\displaystyle\dot H_1&\displaystyle=&\displaystyle\frac{1}{m_p^2}\Bigl(1-
\tanh(t)^2\Bigr)\dot\phi_1,\\[3.7mm]
\displaystyle\dot\phi_1&\displaystyle=&
\displaystyle\frac{\left(18-24m_p^2
+24(m_p^2-1)\tanh(t)^2+6\tanh(t)^4\right)\tanh(t)}
{12m_p^2\left(1-\tanh(t)^2\right)}
 \phi_1-{}\\[2.7mm]&\displaystyle-&\displaystyle
 \frac{\left(3-\tanh(t)^2\right)\tanh(t)}{1- \tanh(t)^2}H_1.\\
\end{array}
\label{equeps}
\end{equation}

System~(\ref{equeps}) has the following solution:
\begin{equation}
\begin{split}
\phi_1(t)&\displaystyle=2m_p^2C_1\left(1-\tanh(t)^2\right)+{}\\
&{}+2m_p^2 C_2\frac{2
J(t)+(\cosh(2t)-1)(\cosh(t))^{2-\frac{1}{m_p^2}}
e^{\left(\frac{1}{2m_p^2(\cosh(2t)+1)}\right)}}{\cosh(2t)+1},\\
H_1(t)&=C_1\left(1-\tanh(t)^2\right)^2-\frac{4m_p^2{C}_2 J(t)}
{(\cosh(2t)+1)^2},\\
\end{split}
\label{phi1H1}
\end{equation}
where
\begin{equation*}
 J(t)=\int_0^t \!\sinh\left(\tau\right) \left(
\cosh(\tau)\right)^{1-1/m_p^2}\left(
2\left(2m_p^2-1\right)\cosh(\tau)^2-1 \right)
e^{\frac{1}{4m_p^2\cosh(\tau)^2}}d\tau.
\end{equation*}

If $C_2=0$, then $H_1(t)$ and $\phi_1(t)$ are bounded for all
$t\in(-\infty,\infty)$ and for all values of $m_p$. In this case the solution
can be presented in the following form:
\begin{equation*}
 \phi(t)=\phi_0((1+\varepsilon C_1)t)+{\cal
 O}(\varepsilon^2),\quad
  H(t)=H_0((1+\varepsilon C_1)t)+{\cal O}(\varepsilon^2).
\end{equation*}

Let us consider the case  $C_2\neq 0$. It is easy to see that if
$m_p^2>1/2$ then $\phi_1(t)$ goes to infinity as
$t\rightarrow\infty$ and, therefore, our solution is not stable. At
$m_p^2=1/2$ we obtain from (\ref{phi1H1}) that
\begin{equation*}
\begin{split}
H_1(t)&=\left(\tanh(t)^2-1\right)^2\left(C_1-C_2J_2\right),\\
\phi_1(t)&=-\left(\tanh(t)^2-1\right)\left(C_1-C_2J_2\right)
-\frac{1}{2}C_2e^{-\tanh(t)^2/2},
\end{split}
\end{equation*}
where $J_2=\int_0^t \!{e^{-\tanh(\tau)^2/2}}\tanh(\tau){d\tau}$. Thus, $H_1$
and $\phi_1$  are bounded functions. For $m_p^2<1/2$, $\phi_1(t)$ and $H_1(t)$
are bounded functions as well. Thus we obtain that our solutions are stable,
the first corrections are bounded, if $m_p^2\leqslant 1/2$. We remind, that
for $m_p^2< 1/2$ points $\phi=\pm 1$ are minima, and  for $m_p^2=1/2$ these
points are inflection ones.

 To study a behavior of solutions with initial
conditions much distinct from the initial conditions of the exact solutions
and to answer the question whether the obtained trajectory is an attractive or
repulsive one, we construct phase portraits for various values of $m_p^2$.

Note, that  $H(t)$ is a real function. It gives a restriction on the maximal
value of $\dot\phi(t)^2$, namely, for all $\phi$ the inequality that
$\dot\phi^2\leqslant 2V(\phi)$ is hold. For example, let $\phi(0)=0$ then only
solutions with $\dot\phi(0) \leqslant 1$ have the physical sense. In  other
words our exact solution starts from this point with the maximal possible
speed.

 We can not find exact solutions to system (\ref{eom12}) with
initial conditions $\phi(0)=0$ and $|\dot\phi(0)|<1$, and thus we
present the results of numeric calculations.

\begin{figure}[h]
\centering \large $\qquad\qquad m_p^2=100$, \hfill  $m_p^2=1$,
\hfill  $m_p^2=7/12\qquad\qquad  $\\[2.7mm]
\includegraphics[width=150mm]{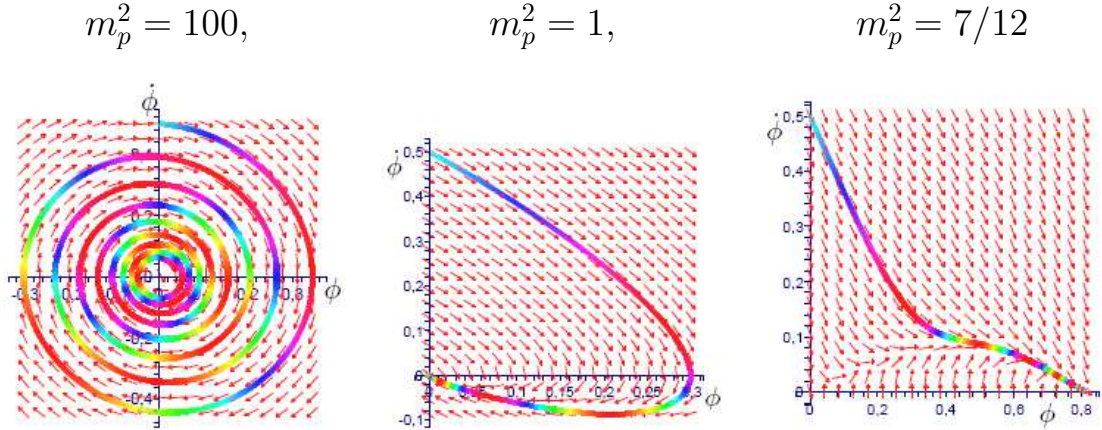}\\
\caption{Phase portraits for different values of $m_p^2>1/2$. Thick
lines strokes depict solutions with the initial condition
$\dot\phi(0)=1/2$.}\label{Phasepo}
\end{figure}

In Fig.~\ref{Phasepo} phase portraits for $m_p>1/2$ are presented, i.e. in the
domain of instability of our solution. One can see on the graphs that
evolution of solutions comes to the end at points of a minimum of flip
potential $-V$, namely at zero for $m_p^2=100$ and $m_p^2=1$ and at the point
$\phi=\sqrt{2/3} \approx 0.816$ for $m_p^2=7/12$ (see. Fig.~\ref{Vflipplot}).

\begin{figure*}[h]
\large\centering $\qquad\qquad m_p^2=1/2$, \hfill $m_p^2=1/4$, \hfill
$m_p^2=1/100\qquad\qquad$\\[2.7mm]
\includegraphics[width=150mm]{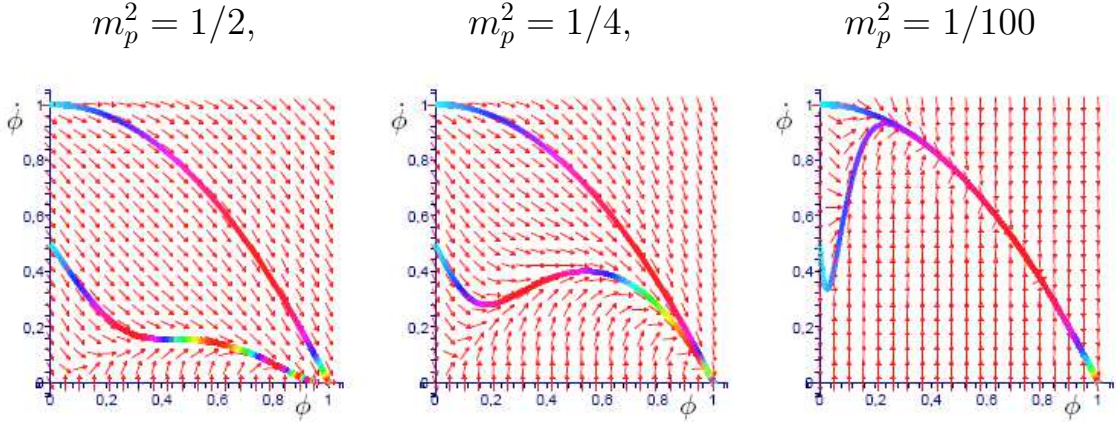}\\
\caption{The phase portraits at $m_p^2\leqslant 1/2$.
Thick lines strokes depict solutions with the initial
conditions $\dot\phi(0)=1/2$ and $\dot\phi(0)=1$.}\label{Phasepo05}
\end{figure*}

Let's consider phase portraits at $m_p^2 \leqslant 1/2 $, presented in
Fig.~\ref{Phasepo05}. In the case $m_p^2=1/2$ when $\phi=\pm 1$ are inflection
points, a solution with the initial speed $\dot\phi(0)= 1/2$, as well as the
exact solution tends to the point $\phi=1$, however coincides with it only in
this point. For $m_p^2=1/4$ and $m_p^2=1/100$ we see, that the numeric
solution is close to the exact solution  for $\phi<1$, and the less $m_p^2$
the less is the value of $\phi$, at which solutions with initial speeds
$\dot\phi(0)=1/2$ and $\dot\phi(0)=1$ get an identical speed.

Thus both behavior of the first corrections and phase portraits show that the
obtained exact solution is stable for $m_p^2\leqslant 1/2$.

\subsection{Variation of the form of the potential}

In this subsection we investigate an influence of variation of the
coefficients in the second summand of the potential (\ref{ourV}) on
the behavior of solutions to system (\ref{eom12}). Consider a
dynamics of system (\ref{eom12}) with the potential
\begin{equation*}
\tilde
V(\phi)=\frac12\left(1-\phi^2\right)^2+\frac{y}{12m_p^2}
\phi^2\left(3-\phi^2\right)^2.
\end{equation*}

Parameter $y$ is introduced so that the points $\phi=\pm 1$ remain
the extrema of the potential. Note, that up to a constant shift and
an overall factor multiplication this is the only possible
modification of an even potential (\ref{ourV}), which leaves the
points $\phi=\pm 1$ to be extrema. Equation (\ref{eomphi}) differs
from the kink equation of motion (\ref{flateom}) by the friction
like term and an extra proportional to $1/m_p^2$ term in the
potential. Parameter $y$ enables us to separate the effects of the
friction and an extra term in the potential and to answer the
question which one provides the stability of solution.

\begin{figure}[h!]
{\small \centering \large $\qquad\qquad\qquad y=0.2$\hfill $y=1$
\hfill
$y=1.8\qquad\qquad\qquad\quad$\\[2.7mm]
\includegraphics[width=140mm]{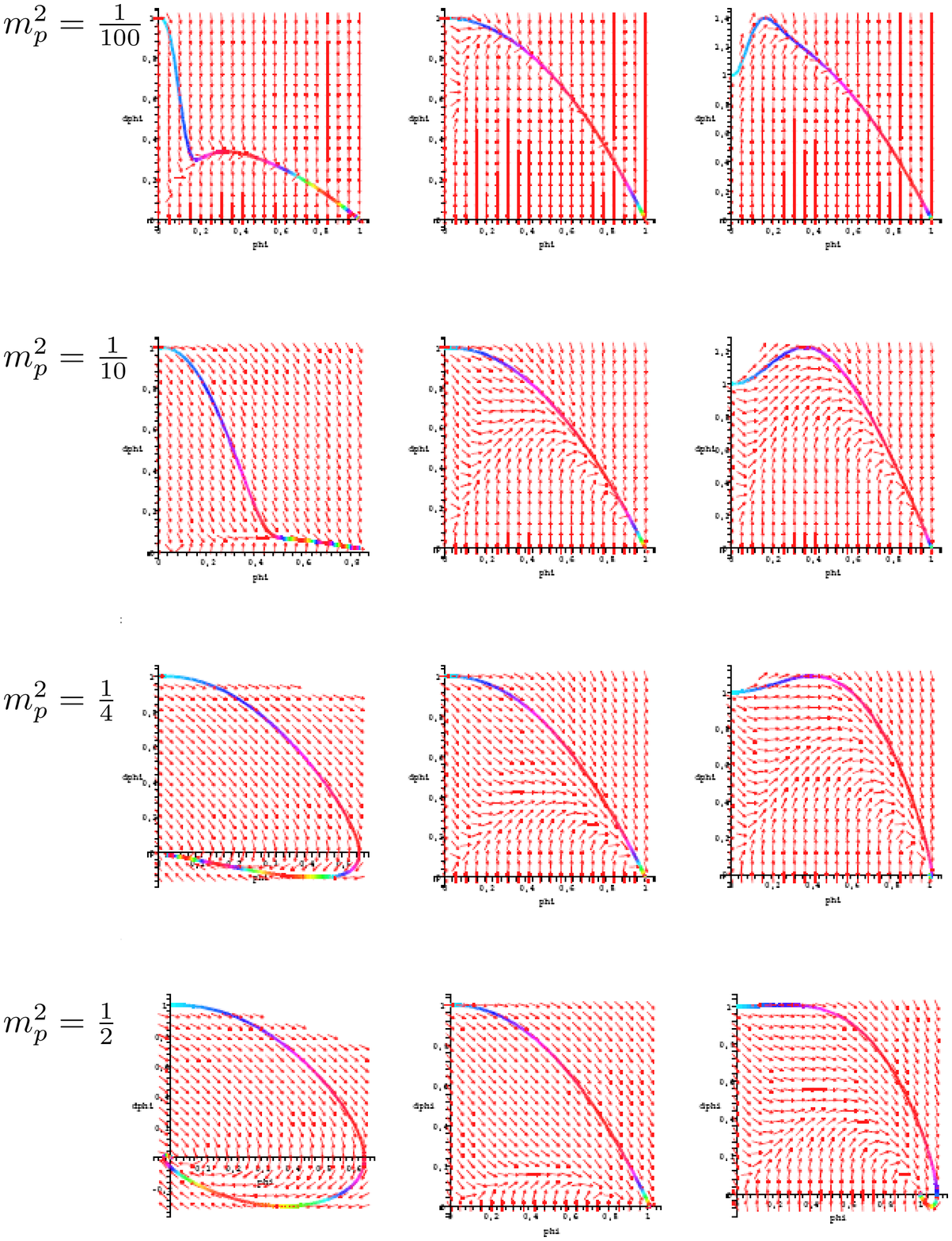}
} \caption{Phase portraits for $m_p^2\leqslant 1/2$ and various $y$.}
\label{phaseportraits1}
\end{figure}

\begin{figure}[h!]
{\centering
\large $\qquad\qquad\quad y=0.2$\hfill $y=1$ \hfill $y=1.8\qquad\qquad\qquad$\\[2.7mm]
\includegraphics[width=127mm]{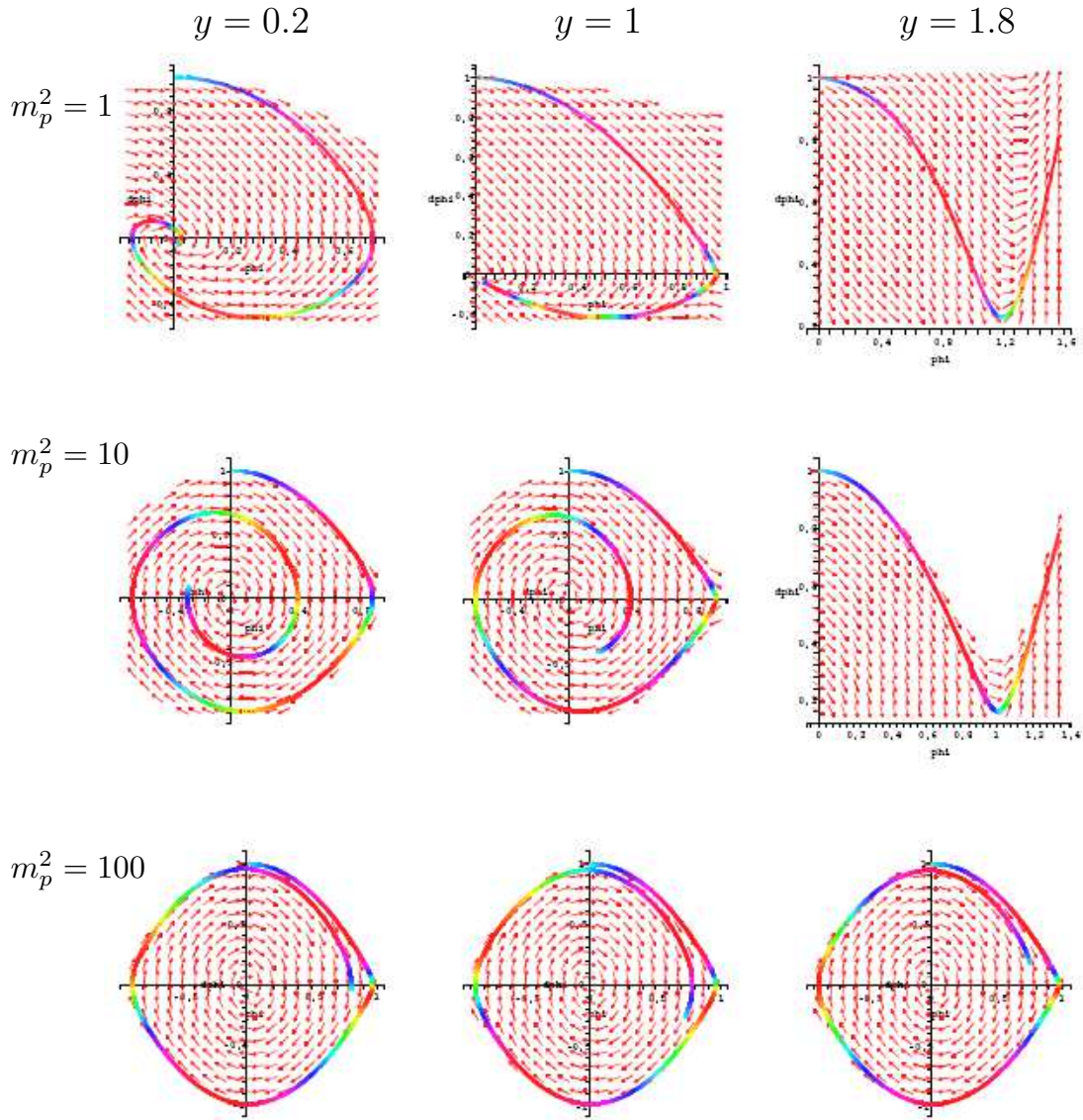}
} \caption{Phase portraits for $m_p^2 > 1/2$ and various $y$.}
\label{phaseportraits2}
\end{figure}

To analyze stability it is convenient to plot phase portraits for different
values of parameters $m_p$ and $y$ and to render them in a table as it is done
in Figs.~\ref{phaseportraits1} and \ref{phaseportraits2}. Guided by the
plotted phase portraits and an analysis performed in Sections 3 and 4, we
conclude the stabilization of solutions is due to an extra term in the
potential. Namely, the obtained solution is attractive one for
$m_p^2/y\leqslant 1/2$ irrespectively the numeric factor in the friction term.

\clearpage

\section{Conclusion}

In the present paper we have performed an analysis of an exactly solvable
model of accelerating expanded Universe dominated by the dark energy. The
state parameter of the model in the consideration is always less than $-1$ and
tends to $-1$ when $t\to\infty$. Such a behavior of the state parameter leads
to an absence of the Big Rip singularity in our model. This kind of
singularity does exist in the model with a constant $w<-1$. Note, that these
properties of the state parameter are kept after including in our model an
interaction with the cold dark matter \cite{AKV-inprogress}. It is shown also
in this paper that our solution is stable with respect to small fluctuations
of initial data when $m_p^2\leqslant 1/2$ and small fluctuations of the form
of the potential. The simplest way to yield an exactly solvable model with $w$
crossing the cosmological constant barrier $w=-1$ consists in an inclusion of
an additional scalar field \cite{AKVtwofields}.


Authors are grateful to A.E. Pukhov and I.V. Volovich for useful discussions.
The work is supported in part by RFBR grant 05-01-00758, I.Ya. and A.K. are
supported in part by INTAS grant 03-51-6346 and Russian President's grant
NSh-2052.2003.1, the work of S.V. is financially supported in part by Russian
President's grant NSh-1685.2003.2 and grant of the scientific program
``Universities of Russia''{} 02.02.503.


\section*{Appendix. Cosmo-phantom versus 5-dimensional gravity models}

Our model as whole and system (\ref{eom12}) in particular are in close analogy
with models considered in brane-world scenarios. Let us compare system
(\ref{eom12}) with the system of the Einstein equations in the model of
gravity interacting with a single scalar field in five-dimensional
space-time~\cite{DeWolfe,Gremm}. The action is
\begin{equation*}
 \tilde S=\int d^4x dr
\sqrt{|\det\,\tilde g_{\mu\nu}|}\left(\frac{\tilde R}{4}-\frac{1}{2}\tilde
g^{\mu\nu}\partial_{\mu}\tilde\phi\partial_{\nu}\tilde\phi- V(
\tilde\phi)\right),
\end{equation*}
 where $\tilde g_{\mu\nu}$ is defined as follows
\begin{equation*}
  d\tilde s^2=e^{2A(r)}(-dx_0^2+dx_1^2+dx_2^2+dx_3^2)+dr^2.
\end{equation*}
Assuming the scalar field depends only on $r$, the independent equations of
motion are
\begin{eqnarray}
&& \tilde{H}'=-\frac{8}{9}(\tilde\phi')^2, \label{eomWolfe1}\\
&&
3\tilde{H}^2=-\frac{16}{9}\left(V(\tilde\phi)-\frac{1}{2}(\tilde\phi')^2\right),
  \label{eomWolfe2}
\end{eqnarray}
where $\tilde H\equiv 4A'/3\equiv \frac{4}{3}\frac{\mathrm{d}A}{\mathrm{d}r}$
and $\tilde\phi'\equiv\frac{\mathrm{d}\tilde\phi}{\mathrm{d}r}$. These
equations correspond to (\ref{eom12}) with $m_p^2=-9/16$. Surely, such a
choice of the $m_p^2$ coupling is not physical, but we see that the method
described in~\cite{DeWolfe} to transform (\ref{eom12}) into the first order
differential equation works in our case as well, i.e. mathematical method,
which simplifies (\ref{eomWolfe1}) and (\ref{eomWolfe2}), does not depend on a
value or sign of~$m_p^2$.

\end{document}